# Recognition and evaluation of constellation diagram using deep learning based on underwater wireless optical communication

ZiHao Zhou[a], WeiPeng Guan[b,*], ShangSheng Wen[c]

([a]School of Information Engineering, South China University of Technology, Guangzhou, Guangdong 510640, China

[b]Department of Information Engineering, The Chinese University of Hong Kong, Shatin, Hong Kong, China

[c]School of Materials Science and Engineering, South China University of Technology, Guangzhou 510640, China)

*Corresponding author: wpguan@ie.cuhk.edu.hk   &   gwpscut@163.com

**Abstract.** In this paper, we proposed a method of constellation diagram recognition and evaluation using deep learning based on underwater wireless optical communication (UWOC). More specifically, an constellation diagram analyzer for UWOC system based on convolutional neural network (CNN) is designed for modulation format recognition (MFR), optical signal noise ratio (OSNR) and phase error estimation. Besides, unsupervised learning is used to excavate a new optimization metric from various factors that affect the quality of underwater channel. The proposed new metric synthesizes several original indexes, which we termed it as multi noise spatial metric (MNSM). The proposed MNSM divides the quality of constellation from high to low into several levels and reflects the quality of UWOC channel. Through the simulation, the constellation diagrams of four widely used M-QAM modulation formats for 16 OSNR values (15dB~30dB) are obtained, with the phase error standard deviations ranging from 0° to 45°. The results show that the accuracy of MFR , the estimation of OSNR and phase noise are 100%, 95% and 98.6% accuracies are achieved respectively. The ablation studies are also carried out in order to analyze the performance of deep learning in the recognition of constellation diagrams.

## 1. Introduction

A growing number of underwater applications, such as oceanography studies and ocean environment monitoring [1,2], are putting forward higher requirements for communication technology. Therefore, as a suitable and efficient transmission solution for manifold underwater applications, UWOC system has recently attracted considerable attention. Compared with the traditional underwater acoustic communication, UWOC is more secure and can provide lower time delay and higher bandwidth [3] which is more compatible with underwater ecosystem. Based on the above advantages, the research on UWOC is also continuously in-depth in recent years. Undeniably, how to measure the quality of optical signal effectively and accurately is one of the most important tasks of in the UWOC research. Eye diagram can reflect the quality of optical signal well. But over the past few years, UWOC based on coherent optical transmission system has adopted the progressive modulation format such as quadrature amplitude modulation (QAM). In this case, the eye diagrams are not effective due to the lack of phase information. On the contrary, the advantages of constellation diagram are fully demonstrated at this time. After M-QAM modulator, the encoded binary bit stream will be divided into in-phase (I) and quadrature (Q) signals. Constellation diagram maps those two signals onto the I-Q plane to express the signals and their relationship intuitively. The constellation diagram has become an important and effective method to measure and evaluate the quality of optical signal of this kind of system since it contains the information of amplitude, phase and many other transmission indexes. The analysis of constellation diagrams is often used to recognize modulation format and analyze various kinds of damages of channel. Furthermore, the shape of constellation diagrams and the distribution characteristics of constellation points can provide clues about the sources and types of interference. However, the traditional analysis of constellation diagrams relies heavily on the professional expertise, which is only suitable for the experienced engineer[4]. And if the analysis is carried out manually, meaning that all the in-phase and quadrature data need to be collected, which will be a huge and inefficient project, not suitable for real-time detection, and it is difficult to get accurate results. Therefore, we urgently require a more advanced constellation analysis method, which can achieve a higher accuracy and speed without human intervention.

On the other hand, as we all know that one of the instructive principles of communication system design is to divide it into several signal processing sub modules, each of which performs a specific function. However, it is not determine that the group composed of individually optimized signal processing module achieve the best possible end-to-end performance. And each independent module has a variety of corresponding local evaluation indicators, try to optimize each of these components according to these different local indicators may lead to unwieldy and computationally complex systems. Furthermore, prior studies have shown that the optical beam is affected by absorption, scattering [5] and ocean turbulence when propagating in seawater. Therefore, under the influence of so many interference factors, in order to ensure the transmission quality of UWOC system, we must optimize the communication system end-to-end according to a comprehensive evaluation metric. But there is no such a evaluation metric in the existing research, and it is quite difficult to extract an end-to-end evaluation metric manually from so many optical performance evaluation indexes.

It will be a rather complex and huge project when the above two problems (constellation analysis and global optimization of UWOC) are dealt with separately by traditional methods. In view of the above two problems, in this paper, we proposed a method of recognition and evaluation of constellation diagram using convolutional neural network (CNN) based on UWOC. The proposed method can realized the recognition of (8QAM, 16QAM, 32QAM, 64QAM) and the estimation of OSNR and phase error in UWOC, which can achieve higher accuracies. What's more, unsupervised learning is used to cluster the OSNR and phase error detected by CNN, and a new global optimization metric called MNSM is obtained, which can be used to design and end-to-end optimize the UWOC system. However, it should be noted that in our first work, CNN is used to identify OSNR and phase error. In order to maintain the correlation between the two work, we only used k-means clustering to fuse these two indexes. But the idea of unsupervised learning is not limited to the integration of these two indicators. We can use this idea to synthesize more indicators to get a more comprehensive evaluation plan. Finally, section 2 and section 3 respectively introduce the theoretical basis of CNN and unsupervised learning. And section 4 focuses on the establishment of simulation system and the analysis of results. The proposed technique has the potential to conduct real-time underwater optical performance monitoring and guide the end-to-end optimization of UWOC.

## 2. Design of the Proposed Constellation Analyzer

CNNs are a category of Neural Networks that have been proven very effective in areas such as image recognition and classification. They are especially suitable for processing data that have a grid-like topology, and image data just meet such structural characteristics. The architecture of the proposed constellation diagram analyzer based on CNN is summarized in Fig.1. The CNN we built for experimental training mainly includes four operations: Convolution, Non Linearity (ReLU), Pooling or Sub Sampling and Classification (Fully Connected Layer). In order to reduce the computational complexity without losing too much pixel information, the colored image is converted into grayscale and downsampled to 64 × 64 for CNN's input data. The first convolutional layer ($Conv1$) filters the 64×64×1 input image with 6 kernels of size 5×5 with a stride of 1 pixels. Then the data of size 60×60×6 are sent into the pooling layer 1($P_1$), whose subsampling region is 2 × 2. The second convolutional layer ($Conv2$) takes as input the output of the first pooling layer ($P_1$) and filters it with 12 kernels of size 5×5. The second pooling layer ($P_2$) still downsamples the input from $Conv2$ in 2 × 2 region. After the second maxpooling, the output of $P_2$ will be expanded into a one-dimensional vector ($FC_1$). The number of neurons in fully-connected layer $FC_1$ and $FC_2$ is 2028 (12 × 13 × 13) and 192. Then, $FC_2$ is fully connected with three output layers ($fc_1, fc_2, fc_3$) respectively. We used softmax as activation function for both $fc_1, fc_2$ and $fc_3$. The number of neurons in the three output layers is 15, 4, 5 where $fc_1$ outputs the probability of each OSNR value, while the $fc_2$ and $fc_3$ layers predict the modulation format and phase error respectively. In the whole CNN training, Adam optimization algorithm is used to adjust the value of neuron parameters to get better convergence and faster training speed. Below, we will describe some of the buliding blocks our network's architecture. Sections 2.1-2.4 are sorted according to the function of each module.

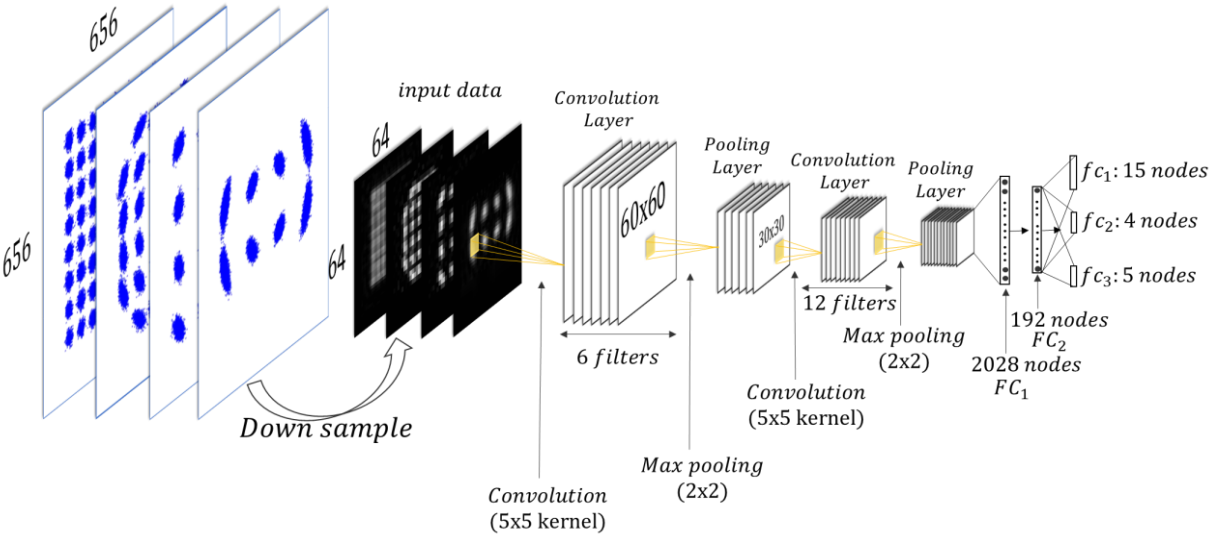

Fig. 1. Architecture of the proposed constellation analyzer based on CNN. Input layer: Gray constellation with pixel size of 64 × 64. Conv1 is composed of 6 kernels of 5×5 , and the stride of convolution is 1 pixel. Conv2: Flters the 30×30×6 input feature maps with 12 kernels of size 5×5 with a stride of 1 pixels. Max pooling layer P1 and P2 sample the input characteristic map through 2x2 region. Fully connected layer 1 ($FC_1$): 2028 nodes are obtained after expanding the feature map. The number of neurons in $FC_2$ is 192. The output layers $fc_1$, $fc_2$ and $fc_3$ perform three tasks respectively: OSNR estimation, modulation format recognition and phase error estimation. Their number of neurons is 15, 4 and 5, respectively.

**2.1 Image Feature Extraction**

CNN derive its name from the "convolution" operator, and it is the introduction of convolution that makes CNN unique. The primary purpose of convolution in case of a CNN is to extract features from the input image, it preserves the spatial relationship between pixels by learning image features using small squares of input data. As we discussed above, every image can be considered as 2D matrix of pixels. In Fig.2. , Take 4 × 4 gray-scale input image as an example, we will demonstrate how convolution works over such a grayscale image.

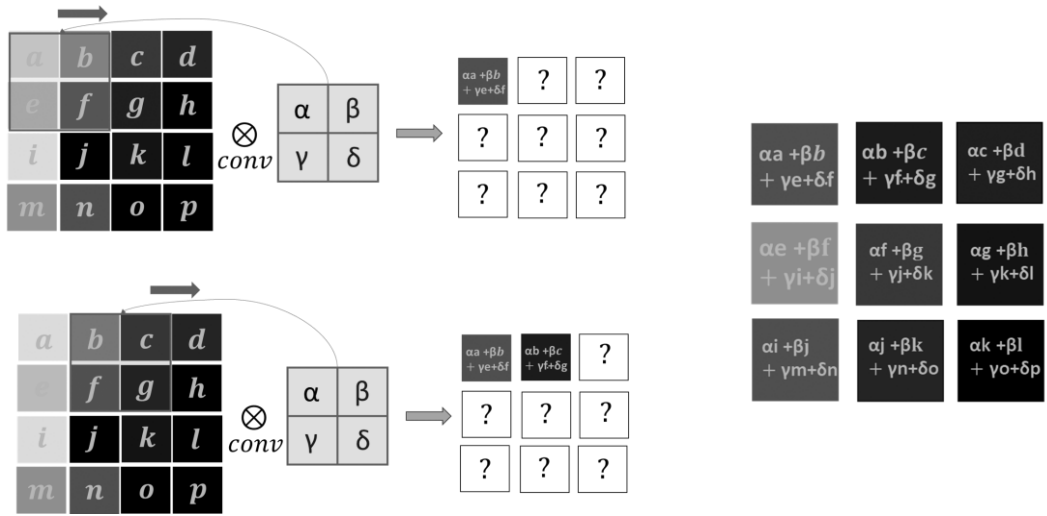

Fig.2. (a) The calculation process of image convolution  (b) The result of conv

As shown in Fig.2.(a), we slide the red kernel over our original gray image by 1 pixel (also called 'stride') and for every position, calculate the product of the elements in the corresponding positions of the overlapped parts and add them together to get a result, which constitutes a single element of the output matrix. Actually, in CNN terminology, we also term the kernel as feature detector. Then it is evident that different values of the feature detector will produce different feature maps for the same input image. It means that different filters can detect different features from a image, for example edges, curves ect, which is helpful for our CNN to analyze and extra the details of the input constellation diagrams. Therefore, in order for CNN to extract enough effective information from the image, we used a variety of convolution kernels.

**2.2 Nonlinear Activation**

An additional operation called ReLU has been used after feature extraction in convolution layer. ReLU is an element wise operation (applied per pixel) and replaces all negative pixel values in the feature map by zero, then the neurons are not activated. It means that only some neurons will be activated at the same time, which makes the network quite sparse, and thus very efficient for computing. This operation performed by the RELU function is referred to in the terminology as unilateral suppression. And the main purpose of ReLU activation function is to introduce non-linearity in our CNN, since most of the real-world data we want our CNN to learn would be non-linear.

**2.3 Down Sampling**

The next step is pooling, pooling layers reduce the dimensionality of each feature map but retains the most important information. In this experiment, we used max pooling. In case of max pooling, we define a spatial neighborhood (for example, a 2×2 window) and take the largest element from the rectified feature map within that window. Fig.3. shows the process of Max pooling, we slide a 2 x 2 window by 2 strides, and take the maximum value of each region. By using max pooling, we can select superior invariant features, improves generalization performance and prevent the model from over-fitting.

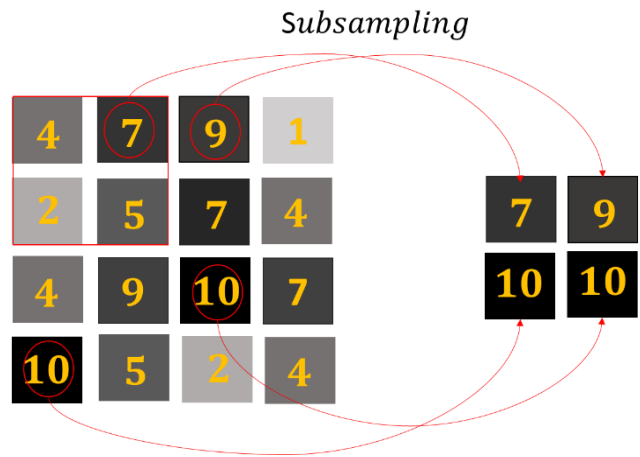

Fig.3. The calculation process of max pool.

**2.4 Feature Integration and Classification**

The Fully Connected layer is a traditional Multi Layer Perceptron that uses a softmax activation function in the output layer . And the term "Fully Connected" implies that every neuron in the previous layer is connected to every neuron on the next layer. Fully connected layer is often used as classifier in CNN. In the case of single task learning, the sum of output probabilities from the fully connected layer is one which is ensured by using the softmax as the activation function in the output layer of the Fully Connected Layer.

# 3. The Generation of MNSM through Unsupervised Learning

If only single evaluation metric is under consideration, the assessment of communication quality will become easier and quantifiable. However, as we said above, the existing single indicators (such as OSNR, phase error, etc.) can only reflect the performance of one aspect of the channel, which is difficult to fully reflect the channel. At the same time, the relationship between these indicators has not been explored in the existing research. In order to carry out the end-to-end optimization of UWOC system, we must synthesize different factors from these properties of constellation response, then extract a new evaluation index. But how to balance these factors also strongly depends on the professional expertise. This paper makes use of the advantages of unsupervised learning algorithm: Using computer instead of manual to realize data annotation and find a comprehensive feature, we termed it as MNSM. K-means clustering algorithm is used to achieve this goal. But it is worth noting that the other work of this paper is to use CNN for constellation analysis. In order to maintain the correlation, we only mined the relationship between OSNR and phase error when we adopted k-means clustering. But the idea of unsupervised learning is not limited to the integration of these two indicators. In our future work, we will consider the integration of more indicators such as error vector magnitude(EVA) and error bit rate (BRE) etc. Next, we will explain the generation of mnsm from the perspective of principle and visualization.

In our experiment, we mainly considered the influence of additive white Gaussian noise and multiplicative phase error on constellation diagrams, we regarded each constellation diagram as a data point, therefore, the job of k-means is to divide M samples into K clusters such that the sum of squared errors for data points within a group is minimized.[9] Let {X(1), X(2)…X(M)} be the set of data points which is to be split into K clusters, each sample can be n-dimensional data. $\mu_1, \mu_2 ... \mu_K$ are K clustering center points[10]. Fig.4. shows the process of a two-dimensional K-means clustering, the horizontal and vertical axes can be regarded as OSNR index and phase error, so X(i) can be written in this coordinate form: (OSNR, phase error).

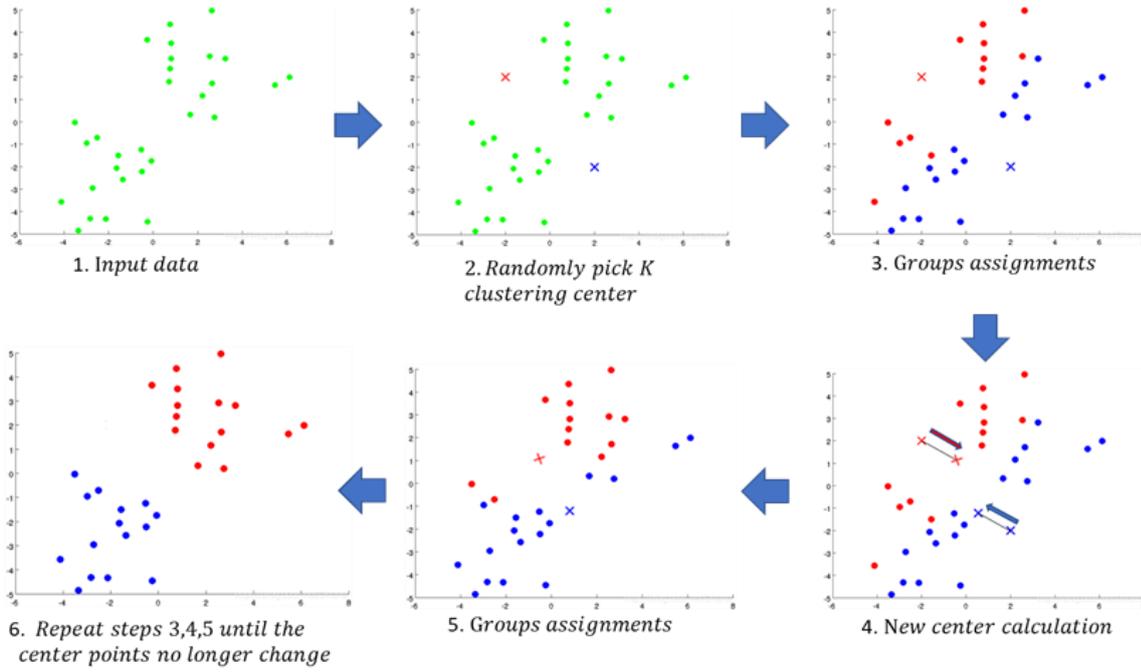

Fig.4. Example to illustarte initialization of K-means algorithm. The data points are shown as dots and cluster centers are depicted as crosses.

Firstly, the algorithm randomly initializes K clustering centers $\mu_1, \mu_2 \ldots \mu_K$. This is followed by two iterative steps. In the first step, each data point X(i) is assigned to the cluster center μ with the minimum Euclidean distance. $C_k$ represents the data set of clustering center with lower case k as the subscript.

$$C_k = \{X(i): \| X(i) - \mu_K \| < \| X(i) - \mu_j \| \ \forall j \in \{1,2 \ldots K\}\{k\}\} \qquad (1)$$

In the next step, the average value of coordinates is calculated for each data point belonging to cluster $C_k$, and this value is taken as the new center point of the cluster, as shown in Eq. (2) .The two steps are repeated iteratively until the cluster centers converge.

$$\mu_k = \sum_{X(i) \in C_k} X(i) \qquad (2)$$

In our experiment, k-means clustering algorithm is used to mine other evaluation metric of these datas. The algorithm finds out the most possible classification (or the optimal solution) from different point classification modes. The $\mu_k$ in Eq (2) will converge to the cluster center point under this optimal classification mode in the continuous iterative calculation. The index of constellation diagrams classification according to this pattern is called MNSM. Figure.5. below shows the exploration process of this optimal classification mode. Note that the leftmost image in Fig.5 represents the effect of initial chaotic distribution of all points on the plane (not drawn to avoid visual confusion), where green points represent randomly initialized clustering centers. The proposed MNSM can guide us to the end-to-end optimization of UWOC system.

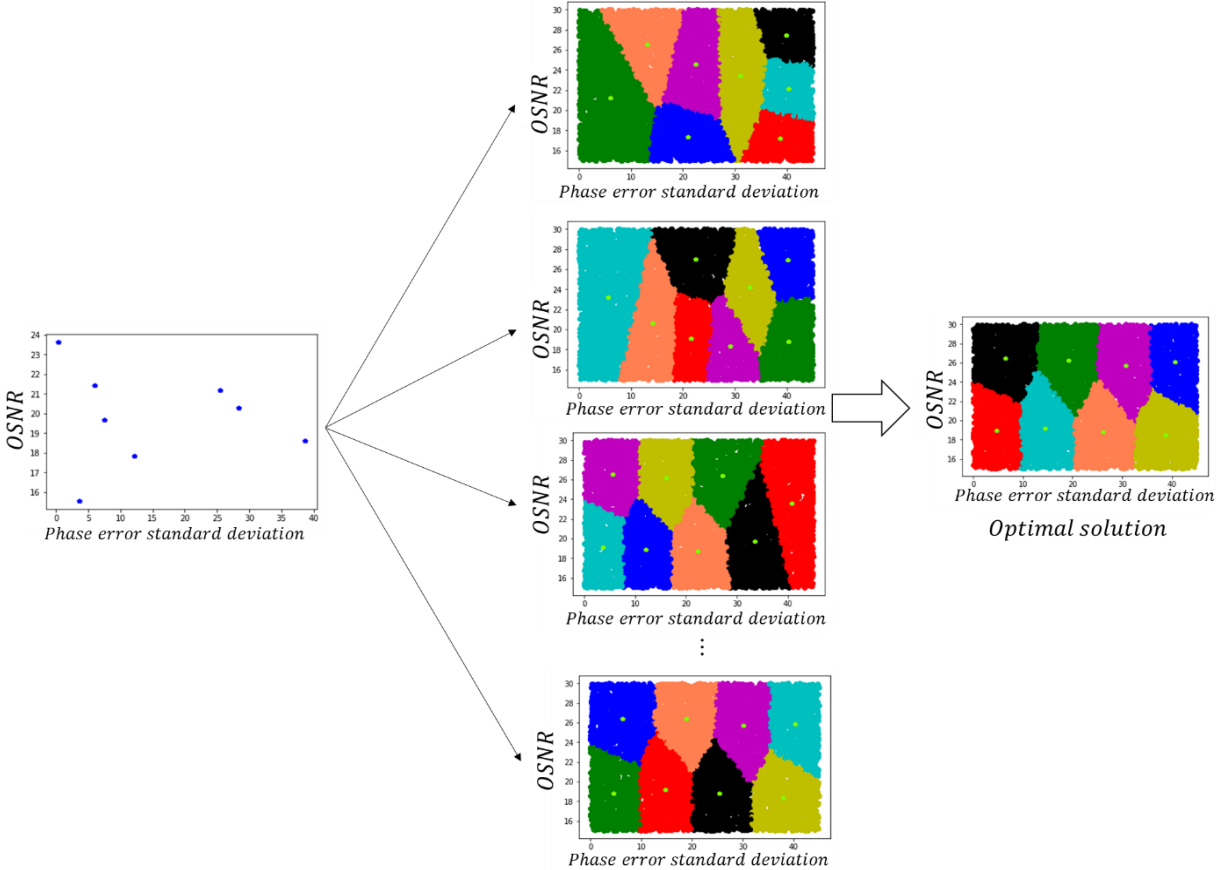

Fig.5. The exploration process of MNSM

## 4. Simulation and Result Analysis

**4.1 Simulation Setup**

Our underwater wireless optical communication system model is shown in Fig.6. First of all, we established the UWOC model (Section 4.2 to 4.4 will provide a detailed introduction), four widely-used signals (8QAM, 16QAM, 32QAM, 64QAM) are modulated by a pseudo-random binary sequence (PRBS) with s symbol number of 1000 and a symbol rate of 155Mb/s. The modulated current input blue LED with cut-off frequency of 6MHz, and the light wave length emitted by LED is $\lambda$ = 450 nm. LED light passed through gamma-gamma channel with different turbulence intensities, which are expressed by scintillation index $\sigma_i^2$. At the reciever, the signal is detected coherently by an optical detector, then after synchronous sampling by two analog-to-digital converters (AD), two electrical digital signals containing the inphase (I) and quadrature (Q) information of four signals are obtained. And we can draw the constellation diagrams of 656 × 656. For the preprocessing stage, the colored images are converted into grayscale and downsampled to 64×64. Then we send the constellation diagrams to CNN for training, where Pytorch$^{TM}$ library is selected as the model of CNN. Finally, mining new evaluation metric from these recognition indexes by k-means clustering.

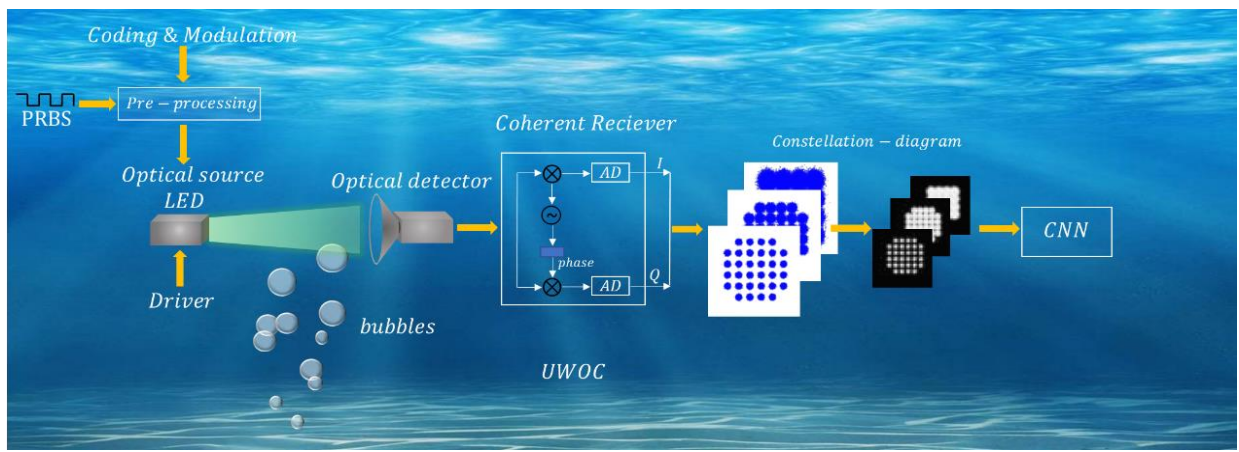

Fig.6. Simulation setup PRBS: pseudo-random binary sequence; AD: analog-to-digital converter

### 4.1.1. Link Budget Analysis for UWOC

In order to obtain the OSNR at the receiver, the received signal power is predicted by modeling the transmitted optical power and the channel path loss while the input-referred noise is estimated by characterizing the noise contribution from each of the receiver building blocks. As shown in Fig.7. the transmitter we used consists of a digital baseband for data coding and signal modulation and an analog front end for driving LEDs. At the receiver side, we had a photodetector (PD), which is used to convert the visible light to electrical current, and then turn the electrical current into voltage and amplified by a transimpedance amplifier (TIA) with a current of $5\times10^{-12}$ A. [11].

A high-pass filter (HPF) is used to remove the low-frequency components of the signal. The reason why we pay attention to the internal structure is that in almost all communication systems, the noise characteristics of the receiver front-end are very important because the incident light power of the front-end is the smallest and it is greatly affected by the noise. The two main noises at the input end of the receiver are the shot noise from the photocurrent and the thermal noise from the receiving circuit[12]. In our simulation, we set the noise bandwidth and noise bandwidth factor as 50MHz and 0.562 respectively. The circuit schematic diagram is shown in Fig.7.

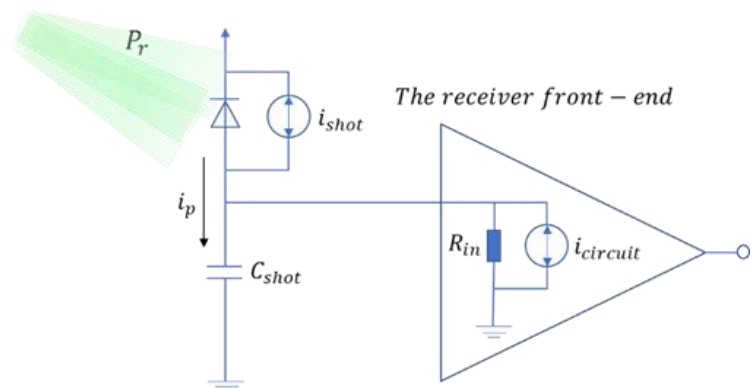

Fig.7. Acceptance schematic diagram of optical receiver front-end

### 4.1.2. UWOC Channel Modeling

In this paper, the description of UWOC channel impulse mainly considers the fading caused by turbulence. We used the gamma-gamma light intensity fluctuation probability distribution proposed by Andrew[13] et al, to model UWOC channel. It is a two parameter model. In such a distribution, the irradiance of the receiver is expressed by the product of two independent random processes[14]. Let $\tilde{h}$ be the channel fading coefficient ($\tilde{h} > 0$), and the probability density function (PDF) of the gamma-gamma distribution is expressed as [15]

$$f_{\tilde{h}}(\tilde{h}) = \frac{2(\alpha\beta)^{\frac{\alpha+\beta}{2}}}{\Gamma(\alpha)\Gamma(\beta)}(\tilde{h})^{\frac{\alpha+\beta}{2}-1}K_{\alpha-\beta}(2\sqrt{\alpha\beta\tilde{h}}) \qquad (3)$$

Where α and β are parameters related to the effective atmospheric conditions, $\Gamma(\cdot)$ is the Gamma function, and $K_n(\cdot)$ is the n th-order modified Bessel function of the second kind. The PDF in Eq. (3) is normalized, so that it satisfies the equality E[$\tilde{h}$]=1. [16] Figure 8. shows the PDF of gamma gamma distribution under several values of α and β.

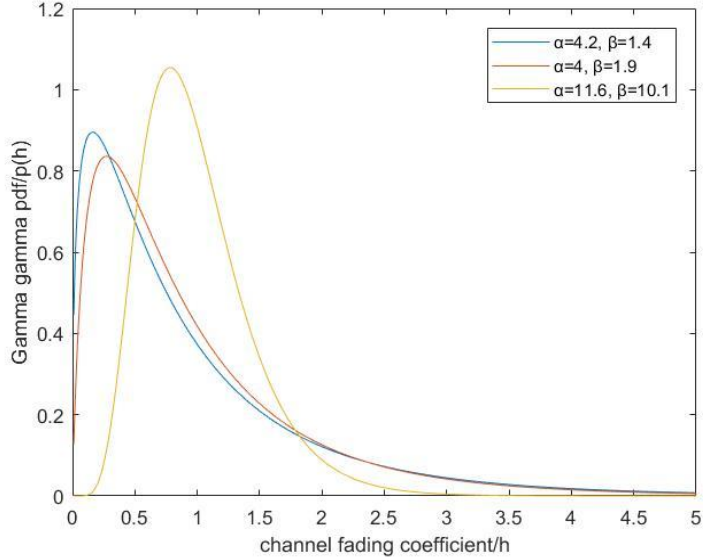

Fig.8. The PDF of gamma gamma distribution under several values of α and β

Furthermore, the scintillation index $\sigma_i^2$ of gamma-gamma model is directly proportional to Rytov variance[17]:

$$\sigma_i^2 = 1.23C_n^2 k^{\frac{7}{6}}L^{\frac{11}{6}} \qquad (4)$$

In the Eq.(4) above, $k = \frac{2\pi}{\lambda}$, $C_n^2$ is a constant related to refractive index. In our experiment, we took $5\times 10^{-14}$. L is the communication distance. We used plane wave in our simulation experiment, when the plane waves travel through the ocean medium, the values of α and β parameters can be calculated by the following formula[17]

$$\alpha = \left[\exp\left(\frac{0.54\sigma_i^2}{\left(1+1.22\sigma_i^{12/5}\right)^{7/6}}\right) - 1\right]^{-1} \qquad (5)$$

$$\beta = \left[\exp\left(\frac{0.509\sigma_i^2}{\left(1+0.69\sigma_i^{12/5}\right)^{5/6}}\right) - 1\right]^{-1} \qquad (6)$$

### 4.1.3. Phase Error of UWOC

In the modulation format, multi-level QAM (M-QAM) is widely used because of its high frequency utilization and good performance, but with the increase of the number of M-QAM signal levels, it is more sensitive to the carrier phase noise, which has a serious impact on the performance of M-QAM, and even the phenomenon of symbol error rate base appears[18]. Phase noise can also cause the rotation of constellation points around the center in the constellation transmitted by M-QAM. Therefore, the detection of phase noise is of great significance to the performance optimization of the system.

In optical communication system, the phase noise mainly includes the phase noise of optical devices, the phase noise caused by atmospheric turbulence and the phase noise of the receiver's demodulator. Our experiment mainly focuses on the phase noise caused by demodulator. It is mainly generated by the random phase jitter of the local oscillator output in the PLL and the phase difference between the signal phase of the receiver. The PDF of phase noise $\varphi$ distribution is expressed as[19]

$$p(\varphi) = \frac{\exp(\cos(\varphi)/\sigma_\varphi^2)}{2\pi I_0(1/\sigma_\varphi^2)}, \quad |\varphi| \leq \pi \tag{7}$$

in which $\sigma_\varphi^2$ is the variance of phase noise, $I_0(\cdot)$ is the zeroth order modified Bessel function of the first kind. In the case of low phase noise deviation, Gaussian distribution can be used to describe the phase noise model. Fig.9. shows the constellation diagrams with phase error standard deviations of 30 ° and 45 ° for OSNR of 25dB.

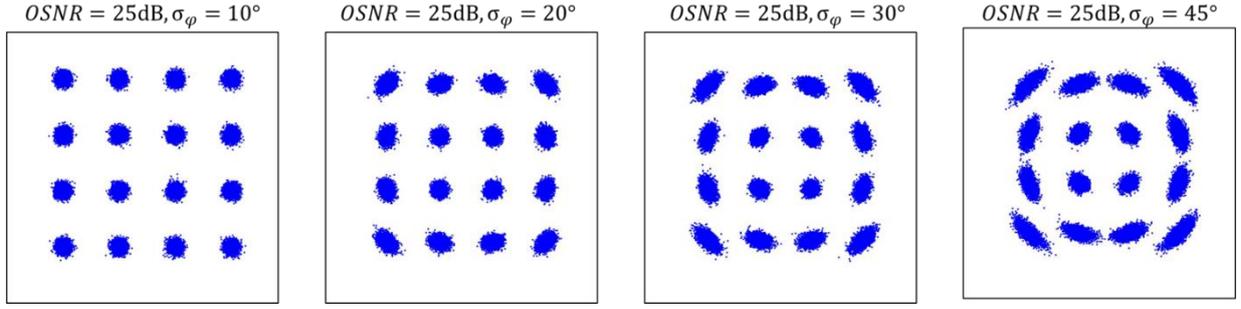

Fig.9. The constellation diagrams with $\sigma_\varphi$ of 10°, 20°, 30° and 45° for OSNR of 25dB.

### 4.2. Results and Analysis

Based on the above system, through the optical wireless link budget analysis, we got 15 different OSNR values from 15dB to 30dB. Each OSNR takes 150 constellation samples, including 30 images with phase noise standard deviation $\sigma_\varphi$ of 0°, 10°, 20°, 30° and 45°. More specifically, each modulation format has 2400 constellation diagram images for 16 OSNR values, so the entire training data set comprises 9600 in total. The RGB constellation diagrams of $656 \times 656$ are obtained through simulation. We downsampled it to $64 \times 64$, and then transformed it into a single channel gray-scale image through color space conversion. We used a series of binary sequences to form vector labels for identifying indicators: Modulation format, OSNR and phase noise. Take the modulation format as an example, in our experiment, we need to identify four modulation formats, so we use 4-bit binary sequence to represent the corresponding modulation formats: 1000 for 8QAM, 0100 for 16QAM, 0010 for 32QAM and 0001 for 64QAM, the position of 1 indicates different modulation formats. Next, we will analyze CNN recognition and the extraction of the new metric MNSM.

### 4.2.1 Analysis of CNN Training

As shown in Fig.10., we measured the accuracy of modulation format recognition, OSNR and phase error estimation every 10 epochs during CNN training. Obviously, as the number of iterations increases, the accuracy of recognition increases correspondingly. As the epochs are small, the accuracy is low, because CNN has not learned enough features. With the epochs grow and the back propagation going on many times, the convolution kernel parameters are gradually optimized. According to the calculation, CNN's training speed is about 7.1s/epoch (Intel Core i7-8500U CPU and 8.00 GB RAM). When the epochs arrive at 250, the recognition accuracy of each metric is close to convergence, and the recognition accuracy of modulation format reaches 100%. Then the accuracy of phase error estimation is about 98%, and the accuracy of OSNR estimation is nearly 95%. Obviously, compared with the estimation of OSNR and phase error, CNN is easier to recognize the modulation format, 100% accuracies are achieved at lower epochs. This is because for different modulation formats, the pixel intensity distribution on the constellation has sharp differences, which is conducive to CNN feature extraction. In the next several parts of this paper, we will focus on the factors that affect the accuracy of OSNR and phase error estimation respectively.

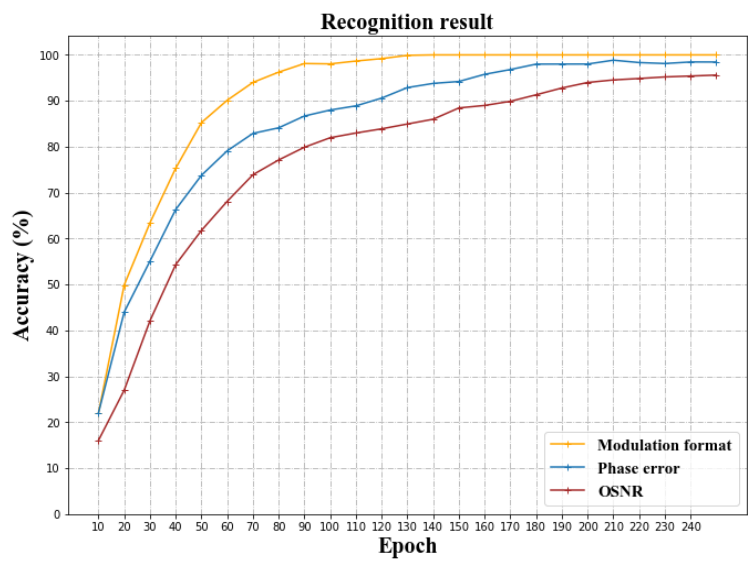

Fig.10. The accuracy of MFR, OSNR and phase error estimation as a function of epochs

### 4.2.2. The Effect of phase Error on OSNR Estimation

We tested the influence of phase error on the accuracy of OSNR estimation in different modulation formats, as shown in Fig.11. We found that in these four modulation formats, most of the misestimated samples of OSNR are distributed in the area with large phase error. Therefore, we took 16QAM and 45 ° phase error standard deviation as an example, and then looked at the OSNR of which part is affected by phase error.

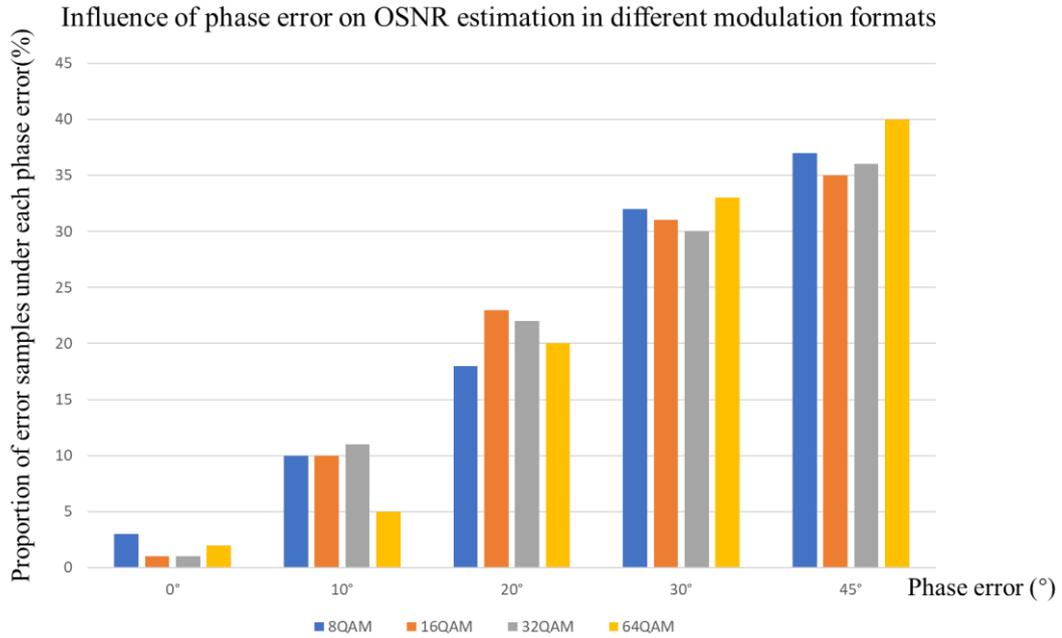

Fig.11. Influence of phase error on OSNR misestimation in different modulation formats

As shown in Fig.12. The result shows that phase noise mainly affects the part of high OSNR. For this reason, we observed the estimation results of OSNR by CNN under the condition of high phase error. Table. 1 shows CNN's estimation of 28dB and 29dB for OSNR when the modulation format is 16QAM and the phase error is 45 ° and 30 ° respectively. Similar results have occurred in other high OSNR cases. Because in high OSNR, if there is no phase noise, the constellation points almost gather on the ideal symbol points, which means the quality of constellation is excellent. However, when the phase error is large, it will cause the rotation of constellation points around the center in the constellation diagram transmitted by M-QAM. And the originally gathered constellation points will become scattered due to the influence of multiplicative noise, which is equivalent to the reduction of OSNR. So in this case, CNN will misjudge OSNR. That is to say, the greater the phase noise, the greater the probability that high OSNR will be misestimated as low OSNR by CNN.

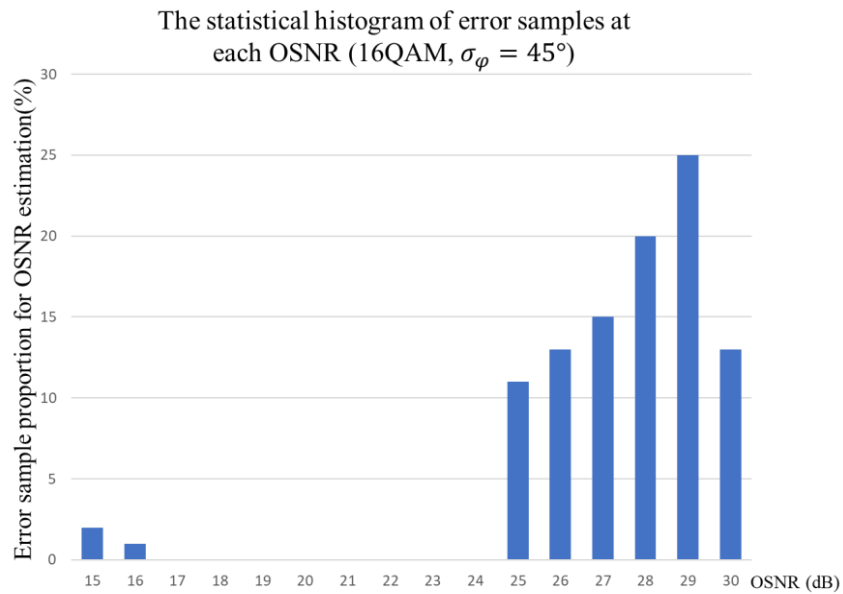

Fig.12. The distribution of OSNR estimation error samples (16QAM, $\sigma_\varphi = 45°$)

Table.1 The influence of high phase error on OSNR estimation

| Original OSNR | Original phase error | Predicted OSNR | Predicted phase error |
|---|---|---|---|
| $29dB$ | 45° | $26dB$ | 45° |
| 28dB | 45° | $24dB$ | 45° |
| $29dB$ | 30° | $28dB$ | 30° |
| 28dB | 30° | $26dB$ | 30° |

### 4.2.3. The Effect of OSNR on Phase Error Estimation

Similarly, we observed the influence of OSNR on phase error estimation in different modulation formats. The statistical results are shown in Fig.13. below.

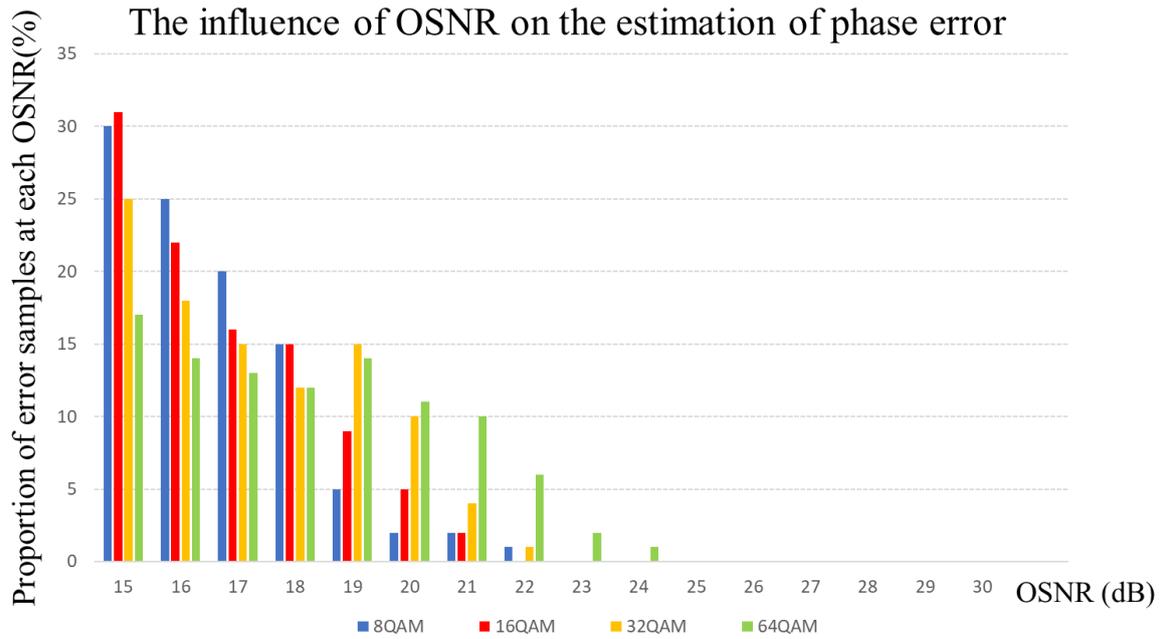

Fig.13. Influence of OSNR on phase error misestimation in different modulation formats

On the contrary, the error samples of phase error estimation are mostly concentrated in the smaller part of OSNR. In Table.2 below, we also observed the estimation of phase error when OSNR is 15dB and 20dB respectively in 16QAM modulation format. Because we know that the influence of OSNR on constellation is that constellation points are distributed in cloud shape with their own symbol points, and the phase error will cause the rotation of constellation points around the center in the constellation diagram. When the OSNR is small, the cloud dispersion of constellation points is obviously stronger than the rotation of constellation points. In other words, in the case of small OSNR, the constellation diagram with high phase error may be misjudged as low OSNR and low phase error by CNN, resulting in error.

**Table.2 The influence of OSNR on phase error estimation**

| *Original Phase error* | *Original OSNR* | *Predicted Phase error* | *Predicted OSNR* |
|---|---|---|---|
| 45° | $15dB$ | 20° | $15dB$ |
| 30° | $15dB$ | 10° | $15dB$ |
| 45° | $20dB$ | 45° | $20dB$ |
| 30° | $20dB$ | 20° | $20dB$ |

## 4.3 Ablation Study

Then we studied the effect of image resolution on CNN recognition accuracy. We transformed the colored constellation diagrams of $656 \times 656$ into the gray scale of $16 \times 16$, $28 \times 28$, $32 \times 32$, $64 \times 64$. Fig.14. (a) shows the accuracies of CNN modulation format recognition under these four different image resolutions. And the accuracies of OSNR and phase error estimation are shown in Fig.14 (b) and Fig.14 (c) respectively. For MFR, the accuracies of these four different resolutions are almost 100% when the epochs exceed 240, which further proves that our CNN constellation analyzer can recognize the modulation format more easily and accurately. For the estimation of OSNR and phase error, it seems that after a period of training, the higher resolution performs the higher estimation accuracy. Because higher resolution provides richer pixel information, which is helpful for CNN to extract features. However, according to prior studies, increasing resolution is not always effective. This is because too high resolution will bring CNN a large number of parameters that need to be adjusted. It is difficult for CNN to adjust the parameters in such a short time, which will lead to the problem of under-fitting. Therefore, in our CNN training, the selected resolution is $64 \times 64$, which can improve the pixel information as much as possible without increasing too many parameters.

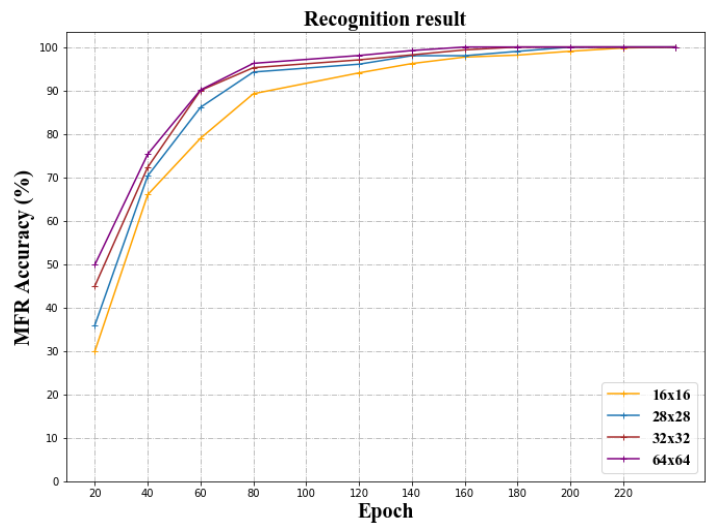

(a)

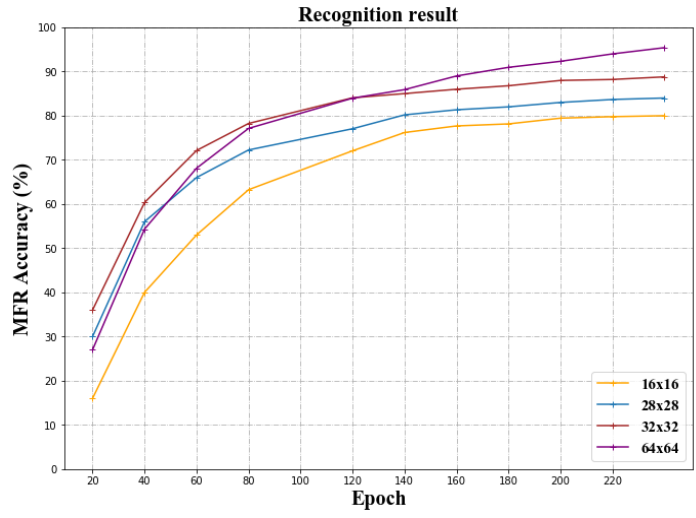

(b)

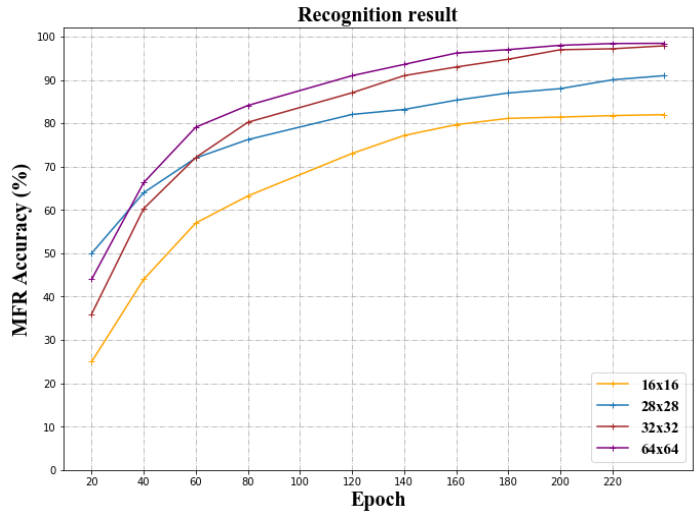

(c)

Fig.14(a). The MFR accuracy at different epochs for input images with different resolutions of $16 \times 16$, $28 \times 28$, $32 \times 32$, $64 \times 64$; (b). The accuracy of OSNR estimation when the image resolution is the four sizes mentioned before. (c). The phase error estimation accuracy with different resolutions mentioned before.

## 4.4 Clustering of the Proposed Evaluation Metrics

After completing the training of CNN analyzer, we mined a new evaluation metric based on unsupervised k-means clustering algorithm. We assumed that the distribution of OSNR is random in range of 15dB to 30dB, and the standard deviation of phase noise $\sigma_\varphi$ is also in the range of 0°~45°. The results of classification of two known metrics of constellation are shown in Fig.15. (a). According to the clustering results of Fig.15. (b), we divided all the original chaotic datas into eight regions, which can represent constellation diagrams of different quality. At the same time, the center coordinates of each class are obtained. The central point in each region based on unsupervised learning, we term it as multi noise spatial metric (MNSM). Because this new metric combines the influence of additive noise and multiplicative noise. And if we continue to fuse more indexes to form high-dimensional data, then this new metric can reflect the spatial relevance of each influence factor. In this way, the new metric we mined can intergrate the performance of many sub modules in a UWOC system, and guide us to optimize the UWOC system with a comprehensive and concise metric instead of dealing with each module separately.

More specifically, MNSM divides the phase noise standard deviation $\sigma_\varphi$ approximately into four intervals from the horizontal direction, OSNR is distinguished by 23dB in vertical direction. For example, we can see that the constellation in the lower left corner belongs to low OSNR case, but the phase error is small, the upper left corner belongs to the constellation with high OSNR and low phase error, so the constellation with high quality belongs to this category, however, when we look at the classification of the lower right corner, not only the phase error is large, but also the OSNR is very low, which belongs to the low quality constellation. In other words, MNSM can be regarded as a constellation scoring mechanism, if the array composed of phase noise standard deviation and OSNR is regarded as a point on the plane, then we can calculate the Euclidean distance between this point and the cluster center (MNSM) to get the classification of constellation quality. What's more, the result of constellation quality judgement we get is not only the influence of a specific module in UWOC system on constellation diagram, but also a combination of the effects of many modules. After classifying the constellation quality, the communication system can be optimized more pertinently.

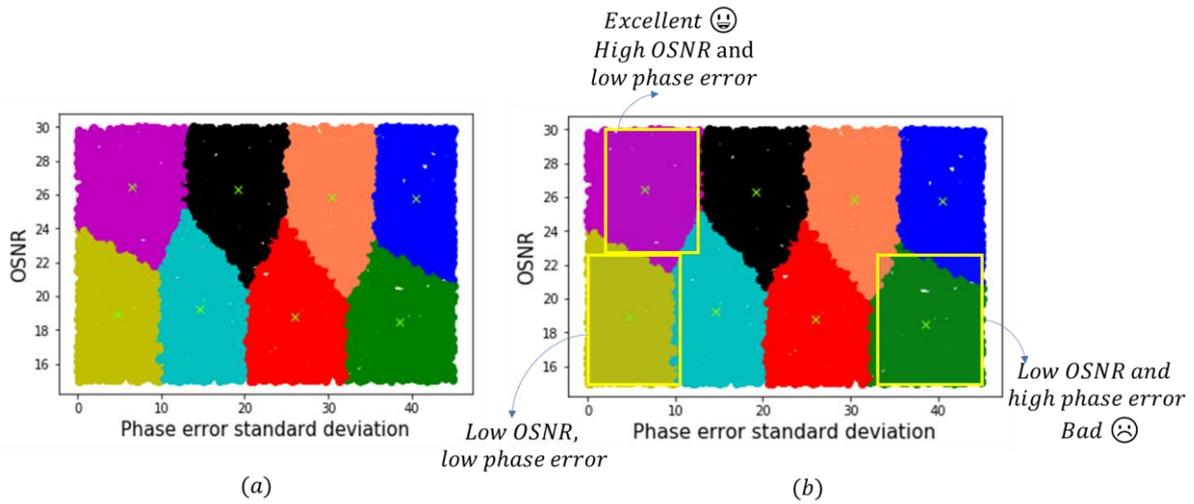

Fig.15. (a) The result graph of new evaluation metric produced by k-means clustering.

(b) Classification of constellation quality by new metric MNSM

Then, for the transmission of an UWOC system, we can predict the modulation format, OSNR and phase error of the transmission system through the constellation processed and transmitted by CNN, and then we can estimate what kind of constellation belongs to by using MNSM mined by k-means. In our future work, we will combine more quality factors to form more complex high-dimensional data to extract new evaluation metrics. Because it seems that clustering with two indicators can also be classified manually, but if we continue to practice this idea of mining more and more complex, that is, higher-dimensional information, it is of great significance.

## 5. Conclusions

In this paper, a method of constellation diagram recognition and quality evaluation using deep learning based on UWOC system is proposed. We simulated the UWOC system, and collected the constellation diagram images of four widely used M-QAM (8QAM, 16QAM, 32QAM, 64QAM) modulation formats in the OSNR range(15dB~30dB) and four different phase errors standard deviations. For MFR, 100% accuracies are achieved, and the estimation accuracies of OSNR and phase error are 95% and 98.6% respectively. And we successfully used k-means to extract a whole new evaluation metric called MNSM, which fuses the original characteristics of different indexes. It is significant for the end-to-end optimization of UWOC. We can also use k-means clustering to mine new comprehensive constellation evaluation metrics from more indexes (such as EVM, BER, linear or nonlinear influences) in our future work. We believe that this technology has the potential to be embedded in the test instrument to realize the intelligent signal analysis and evaluation, thus guiding the global optimization of UWOC system and ensuring the stable operation of optical network.